\documentclass[epj,bio,mat,itm,e3s,shs]{webofc}
\usepackage[varg]{txfonts}   % Web of Conferences font
\usepackage{listings}
\usepackage{hyperref}
\begin{document}
\title{Little Bear's Pulsating Stars: Variable Star Census of dSph UMi
Galaxy}

\author{\firstname{K.} \lastname{Kinemuchi}\inst{1,2}\fnsep\thanks{\href{mailto:kinemuchi@apo.nmsu.edu}{\tt kinemuchi@apo.nmsu.edu}} \and
        \firstname{E.}
        \lastname{Jeffery}\inst{3}\fnsep\thanks{\href{mailto:ejeffery@byu.edu}{\tt ejeffery@byu.edu}}
          \and
\firstname{C.}
        \lastname{Kuehn}\inst{4}\fnsep\thanks{\href{mailto:charles.kuehn@unco.edu}{\tt
          charles.kuehn@unco.edu}}
          \and
\firstname{K.}
        \lastname{Grabowski}\inst{1,2}\fnsep\thanks{\href{mailto:kgrabowski@apo.nmsu.edu}{\tt
          kgrabowski@apo.nmsu.edu}}
          \and
\firstname{J.}
        \lastname{Nemec}\inst{5}\fnsep\thanks{\href{mailto:jmn@isr.bc.ca}{\tt
          jmn@isr.bc.ca}}
        % etc.
}

\institute{Apache Point Observatory, P.O. Box 59, Sunspot, NM 88349, USA
\and
New Mexico State University, Department of Astronomy, Las Cruces, NM 88001, USA           
\and
Brigham Young University, Provo, UT, USA
\and
University of Northern Colorado, Greeley, CO, USA
\and 
Camosun College, Victoria, B.C. Canada
          }

\abstract{
Recent observations and a photometric search for variable stars in the
Ursa Minor dwarf spheroidal galaxy (UMi dSph) are presented. Our
observations were taken at Apache Point Observatory in 2014 and 2016
using the 0.5m ARCSAT telescope and the West Mountain Observatory 0.9m
telescope of Brigham Young University in 2016. Previously known RR
Lyrae stars in our field of view of the UMi dSph are identified, and we
also catalog new variable star candidates. Tentative classifications are
given for some of the new variable stars. We have conducted period
searches with the data collected with the WMO telescope.   Our
ultimate goal is to create an updated catalog of variable stars in the
UMi dSph and to compare the RR Lyrae stellar characteristics to other
RR Lyrae stars found in the Local Group dSph galaxies. 

}

\maketitle

\section{Introduction}

The Ursa Minor dwarf spheroidal galaxy (UMi dSph) is a Local Group
member with a rich variable star population. RR Lyrae (RRL), anomalous
Cepheids (AC), and eclipsing binary stars have been found and
identified \cite{Nemec:88,Kholopov:71,Vanagt:67}.  Two different
epochs of star formation are indicated via the RRL (older population)
and the ACs (intermediate age).  The UMi dSph is
a metal-poor galaxy, and from the analysis of the RRL,
it has been classified as an Oosterhoff-II (OoII) type object.  UMi is
one a handful of known dwarf galaxies with a clear Oosterhoff
classification.

The goals of this project are to revisit UMi dSph galaxy and
reinvestigate the variable star population.  We present our variable
star census based on the northeast portion of the galaxy.  With our
photometric data, we wil re-evaluate and recalculate many of the
stellar characteristics, in particular the pulsational characteristics.

\section{Data acquisition and reduction}

Data sets were collected at the 0.5m Astrophysical Research Consortium
Small Aperture Telescope (ARCSAT) at Apache Point Observatory and at
the 0.9m telescope at the West Mountain Observatory (WMO) of Brigham Young University.
Due to the field coverage of the telescopes, we focused our survey on
the northeast portion of the galaxy, centered at $\alpha=$15:09:11.34 and
$\delta=$67:15:51.7.  With the ARCSAT telescope, we obtained $\sim100$
epochs, and at WMO, over 200 epochs were observed.

Standard data reduction techniques were implemented, and to
standardize the magnitudes, Landolt Standards \cite{Landolt:92} were
observed.  Photometry was performed using DAOPHOT/ALLSTAR packages \cite{Stetson:87}, and period analysis was done primarily with
the dataset collected at WMO.  Period solutions were
obtained using the Supersmoother program \cite{Reimann:94}, which uses
a running linear regression algorithm.  For amplitude, we used a basic
measurement of the difference between maximum and minimum, omitting
outliers 2.5 $\sigma$ from the mean.  

\section{RR Lyrae results}

From the 543 objects identified and photometered, we find 22 RR Lyrae
candidates.  Cross-matching with the finding charts and variable star
catalog of \cite{Nemec:88}, we find 12 ab-type and 10 c-type RRL
stars.  We additionally find 2 anomalous Cepheids, V6 and V11 (using
Nemec's IDs). In Table \ref{stars}, we provide our identified variable
stars' ID, equatorial coordinates (as determined from astrometry.net
\cite{Lang:10}), mean $V$ magnitude, period solution, and variable star type.
The star ID is the same as in \cite{Nemec:88}, unless it is denoted
with an asterisk, then it is a newly identified star.  
%Figure
%\ref{ltc} shows the phased light curves of the variable stars listed
%in Table \ref{stars}.

\begin{table}
\centering
\caption{Identified variable stars in the north portion of UMi dSph.}
\begin{tabular}{cccccc}
\hline
Star & $\alpha_{2000}$ (deg) & $\delta_{2000}$ (deg) & $V$ & Period (d) & type \\\hline
V5 & 227.48149 & 67.474992 &19.69 & 0.766705 & ab \\
V6 & 227.49143  & 67.464966   &18.17 & 0.725528 & AC \\
V7 & 227.44961 & 67.454614 &  19.61 & 0.690328 & ab \\
V8 & 227.294 & 67.451651 & 19.70 & 0.553879 & ab \\
V9 & 227.53442 & 67.419453 & 19.59 & 0.356460 & c? \\
V10 & 227.58564 & 67.405209 & 19.69 & 0.617496 & ab \\
V11 & 227.63938 & 67.417687 & 18.88 & 0.673431 & AC \\
V12 & 227.63475 & 67.395875 & 19.65 & 0.771735 & ab \\
V13 & 227.46745 & 67.403259 & 19.66 & 0.646050 & ab \\
V19 & 227.53903 & 67.357643 & 19.67 & 0.341565 & c \\
V24 & 227.54186 & 67.310734 & 19.83 & 0.301081 & c? \\
V28 & 227.6731 & 67.278736 & 19.70 & 0.308163 & c? \\
V29 & 227.62463 & 67.252691 & 19.73 & 0.409747 & c? \\
V41 & 227.46443 & 67.232054 & 19.84 & 0.490488 & ab \\
V42 & 227.59471 & 67.209778 & 19.74 & 0.646473 & ab \\
V45 & 227.35121 & 67.207436 & 19.02 & 0.216662 & ? \\
V48 & 227.18208 & 67.218741 & 19.69 & 0.687258 & ab \\
V49 & 227.18687 & 67.212797 & 19.73 & 0.415464 & c? \\
V78 & 227.29384 & 67.38137 & 19.69 & 0.372640 & c? \\
V79 & 227.38367 & 67.328828 & 19.70 & 0.313023 & c? \\
V80 & 227.41875 & 67.3245 & 18.74 & 0.512738 & ? \\
v1540* & 227.88001 & 67.426652 & 19.79 & 0.714490 & ab \\
v1678* & 227.19679  & 67.413053 & 19.75 & 0.336900 & c? \\
v1895* & 227.09781  & 67.386081 & 19.45 & 0.490887 & ab \\
v2078* & 227.83088  & 67.361948 & 19.73 & 0.719031 & ab \\
v3176* & 227.32619 & 67.236128 &19.81 & 0.327060 & c? \\
\hline
\end{tabular}
\label{stars}
\end{table}
%
%\begin{figure*}
%\centering
%\includegraphics[width=10cm,clip,angle=90]{figure1.ps}
%\caption{Phased light curves of RR Lyrae and
%  anomalous Cepheid stars.  Periods (days) are in the upper right
%  corners. }
%\label{ltc}      
%\end{figure*}

%

\section{Curious case of V80}

The variable star identified as V80 in \cite{Nemec:88,
  Kholopov:71,Vanagt:67} has an unclear classification.  This star's
variability appears real in our datasets from 2016.  We find the
amplitude of this star to be closer to the published value in
\cite{Nemec:88}, but the periodicity is longer than in any of the
published literature.  The period solution for V80 was 0.512738 days.
We cannot conclusively classify this variable star at this time.
Applying frequency analysis will help understand the nature.

\section{Future Work}
Additional data are expected to be collected in March 2017 and will
cover the southern regions of UMi dSph that has been missed.  With the
fairly complete coverage of the pulsation cycle to many of the RRab
stars, we will calculate the [Fe/H] metallicities through Fourier
decomposition parameters.  This work will allow us to obtain an
independent measurement of the dSph galaxy's metallicity.  

\begin{acknowledgement} 
\noindent\vskip 0.2cm
\noindent {\em Acknowledgments}:  K.K. would like to thank Suzanne
Hawley, director of ARCSAT for the generous time granted for this project. Professional development funds through the Astrophysical Research Consortium allowed K.K. to present this work at this conference and is gratefully acknowledged.
\end{acknowledgement}

% BibTeX or Biber users please use (the style is already called in the class, ensure that the "woc.bst" style is in your local directory)
% \bibliography{woc,ref}

\end{document}